\documentclass[prb,preprint,showpacs,preprintnumbers,amsmath,amssymb]{revtex4}

\usepackage{graphicx}
\usepackage{dcolumn}
\usepackage{bm}
\usepackage{amsmath}
\usepackage[mathscr]{eucal}

\usepackage{color}

\begin{document}

\title{Localized defect states in MoS$_2$ monolayers: electronic and optical properties}
\author{J. Kunstmann$^1$}
\email{jens.kunstmann@tu-dresden.de}
\author{T.B. Wendumu$^{1,2}$}
\author{G. Seifert$^1$}
\affiliation{$^1$Theoretische Chemie, Technische Universit\"{a}t Dresden, 01062 Dresden, Germany}
\affiliation{$^2$Max-Planck Institut f\"{u}r Physik komplexer Systeme, N\"{o}thnitzer Str. 38, 01187 Dresden, Germany}

\begin{abstract}
Defects usually play an important role in tuning and modifying various properties of semiconducting or insulating materials. Therefore we study the impact of point and line defects on the electronic structure and optical properties of  MoS$_2$ monolayers using density-functional methods. 
The different types of defects form electronic states that are spatially localized on the defect. 
The strongly localized nature is reflected in weak electronic interactions between individual point or line defect and a weak dependence of the defect formation energy on the defect concentration or line defect separation. 
In the electronic energy spectrum the defect states occur as deep levels in the band gap, as shallow levels very close to the band edges, as well as levels in-between the bulk states. Due to their strongly localized nature, all states of point defects are sharply peaked in energy. Periodic line defects form nearly dispersionless one-dimensional band structures and the related spectral features are also strongly  peaked. 
The electronic structure of the monolayer system is quite robust and it is well preserved for point defect  concentrations of up to 6\%.
The impact of point defects on the optical absorption for concentrations of 1\% and below is found to be very small. For higher defect concentrations molybdenum vacancies were found to quench the overall absorption and sulfur defects lead to sharp absorption peaks below the absorption edge of the ideal monolayer. 
For line defects, we did not find a considerable impact on the absorption spectrum.
These results support recent experiments on defective transition metal chalcogenides.
\end{abstract}
\maketitle

\section{Introduction}

Defects usually play an important role in tuning and modifying various properties of semiconducting or insulating materials. 
Substitutional defects in semiconductors are employed to increase the electrical conductivity of the material. The defects create impurity states (in-gap states) in the band gap near the valence (p-type) or conduction (n-type) band edge that, when thermally populated or depopulated, create free charge carriers and thus enhance the conductivity.
In ionic crystals certain types of defects form luminescent centers, such as color centers (electrons trapped in anion vacancy sites) or paramagnetic impurities (d- or f-element cations). The electronic states related to the localized electrons of the impurities form in-gap states within the large intrinsic band gap of ionic crystals. The localized states of color centers and d-element impurities  couple strongly to the continuum of phonon states and form broad vibronic bands in the absorption or emission spectra. The resulting optical properties are observed as the color of gemstones (e.g. ruby, emerald) and find application in solid-state lasers, white LEDs, luminescent light bulbs or phosphors.
Defects in crystals are also of interest as potential single photon emitters. 
Individual photons can be emitted from quantum systems with localized states such as cold atoms, molecules, semiconductor quantum dots or color centers \cite{Lounis2005a}. 

With the recent rise of 2D materials there has also been a considerable research interest in transition metal dichalcogenides (TMDs) monolayer structures such as molybdenum disulphide (MoS$_2$). 
A MoS$_2$ monolayer (ML) consists of two atomic layers of close-packed S atoms separated by one close packed Mo atomic layer \cite{Chhowalla2013} and it is a semiconductor with an direct optical band gap of 1.9 eV, \cite{Splendiani2010,Mak2010} 
whereas its bulk counterpart has an indirect optical band gap of 1.3 eV \cite{Kam1982a}. Compared to the bulk a ML exhibits stronger photoluminescence and reduced screening leads to strong excitonic effects. \cite{Splendiani2010,Mak2010} 
Because of such advantageous optical and electronic properties, MoS$_2$ is believed to be a promising building block for future applications in nanoelectronics and optoelectronics \cite{Radisavljevic2011,Wang2012a}.

Combining the interesting physics of defects with the unique properties of MoS$_2$ seems very promising and various experimental studies of point defects have been reported \cite{Komsa2013b,Zhou2013,Qiu2013b,Hong2015}.  Among many other insights, these works highlight the importance and abundance of sulfur vacancy defects. Theoretical studies of various point defects \cite{Yuan2014a,Noh2014,Komsa2015,Haldar2015,Pandey2016} consistently explain this with a low formation energy of this defect.
Sulfur vacancy defects are currently believed to be the main reason for the low mobility, observed in back gated field effect transistors using MoS$_2$, grown by chemical vapor deposition.\cite{Pandey2016}
Line defects and grain boundaries (GB) within ML-MoS$_2$ have also been studied experimentally  \cite{Komsa2013b,Najmaei2013,Zhou2013,VanderZande2013,Tongay2013} and theoretically  \cite{Zou2012,Enyashin2013,Ghorbani-Asl2013a,Zou2015}. Both point and line defects introduce in-gap states in semiconducting ML-MoS$_2$. Due to their one-dimensional nature, certain line defects are believed to form metallic wires \cite{Liu2014d,Zou2015,Gibertini2015}. 
Similarly, the edge states of triangular MoS$_2$ platelets are metallic-like \cite{Bollinger2001} and exhibit bright photoluminescence, and recent theoretically studies \cite{Wendumu2014,Joswig2015} explain this result. 
%
Recently, the emission of single photons from WSe$_2$ MLs was reported. \cite{Srivastava2014a,He2015,Koperski2015,Chakraborty2015,Kumar2015} All authors attribute the single photon emission to localized states caused by intrinsic defects. However the detailed nature of these defect states remains to be determined.

In view of these interesting results, this paper aims at a deeper understanding of electronic and optical properties of defective MoS$_2$ ML. 
Because the localized nature of defect states is crucial for single-photon emission, this paper focuses on the degree of localization and studies point defects as function of the defect concentration and 8-4 type line defects as function of their separation length $d$. 
Analysis of highly defective MoS$_2$ samples in the transition electron microscope determined the area density of defects to be as high as $10^{13}$ cm$^{-2}$.\cite{Qiu2013b} This corresponds to defect concentrations of the order 1\% (see Table \ref{tab:concentration}).
The concentration dependence of the optical and transport properties of random point defects in the limit of small concentration (1\% and smaller) were studied within an empirical tight-binding approach by Yuan et al.\cite{Yuan2014a}
Here we are mostly interested in the localization of individual defects and therefore study larger concentrations.


Our results for sulfur mono and divacancies and Mo and MoS$_2$ vacancies and mirror and tilt grain boundaries (line defects) show that defects states in MoS$_2$ ML are strongly localized on the defect. 
The strongly localized nature is reflected in weak electronic interactions between individual point or line defect and a weak dependence of the defect formation energy on the defect concentration or line defect separation. 
Most significantly, these point and line defects create sharp, characteristic peaks within the band gap, but the electronic bulk properties of the ML are robust for defect concentrations of up to about 6\%.
The impact of point defects on the optical absorptions for concentrations of 1\% and below is found to be very small. Similarly, the considered line defects are found to have a almost no impact on the absorption spectrum.



%

\section{Computational Details}
For all electronic structure calculations and structural optimization of systems with point defects, we have utilized the density-functional tight-binding (DFTB) method because it allows to study large systems. \cite{Seifert1986,Porezag1995}
DFTB is based on the density functional theory of Hohenberg and Kohn \cite{Hohenberg1964} in the formulation of Kohn and Sham \cite{Kohn1965}. The single-particle Kohn-Sham eigenfunctions 
are expanded with a set of localized atom-centered basis functions. 
These functions are determined by self-consistent LDA density functional calculations of isolated atoms employing a large set of Slater-type basis functions.
The effective one-electron potential in the Kohn-Sham Hamiltonian is approximated as a superposition of the potentials of neutral atoms. Additionally, only one- and two-center integrals are calculated to set up the Hamilton matrix. We have taken a minimal valence basis, including the 5s, 5p, and 4d orbitals for molybdenum, and the 3s and 3p orbitals for sulfur. States below these levels were treated within a frozen-core approximation.

Moreover, we have used time-dependent density functional response theory within the DFTB formulation (TD-DFRT-TB) for the calculations of all excitation spectra \cite{Niehaus2001}. This is also referred to a linear response theory. \cite{Casida1996}
To obtain the excitation energies, the coupling matrix, which gives the response of the potential with respect to a change in the electron density, has to be built. In our scheme, we approximate the coupling matrix in the so-called $\gamma$-approximation, \cite{Niehaus2001,Frauenheim2002} which allows for an efficient calculation of the excitation energies and the required oscillator strengths within the 
dipole approximation. The $\gamma$-approximation is based on the adiabatic LDA approximation  which does not include electron-hole interactions. Therefore excitonic effects are not described. The proper description of excitonic effects goes well beyond the scope of this work and is currently only possible for systems with only a few atoms per unit cell.\cite{Qiu2013a} The TD-DFRT-TB method, however, allows to study systems with thousands of atoms.

DFTB and TD-DFRT-TB  calculations were performed with the deMon computer code\cite{Koster2011} using periodic boundary conditions. The k-space was sampled at the $\Gamma$-point,
only. Therefore all calculations were done in sufficiently large supercells.

The line defects were structurally optimized with density functional theory (DFT) calculations using the Vienna Ab-initio Simulation Package (VASP).\cite{Kresse1996,Kresse1996a}
These optimization runs were performed with the PBE generalized gradient approximation for the exchange-correlation functional,\cite{Perdew1996} employing the PAW method \cite{Blochl1994a}
and a plane-wave basis set with a kinetic energy cutoff of 364 eV. For the k-point sampling of the Brillouin zone, $\Gamma$-point centered grids and an in-plane sampling
density of 0.1/{\AA}$^2$ were used. The k-space integration was carried out with a Gaussian smearing width of 0.05 eV for all calculations.

In all calculations we simulate infinite 2D layers and use periodic boundary conditions. The unit cells were built with at least 14 {\AA} separation between replicas in the perpendicular direction to achieve negligible interaction. All systems were fully structurally optimized until all inter-atomic forces were below 0.01 eV/{\AA}.\footnote{Mind that DFT and DFTB find slightly different Mo-S equilibrium bond lengths of 2.41 {\AA} and 2.45 {\AA}, respectively. In the electronic properties these differences are reflected in different band gaps of systems with point defect (optimized with DFTB) and line defects (optimized with DFT)} 
Spin-orbit interactions (SOI) in ML-MoS$_2$ induce a small band splitting of ca.~0.15 eV near the valence band maximum but they have negligible influence on the geometry, heat of formation, and the localization of in-gap states.\cite{Molina-Sanchez2013} 
Therefore we neglected SOI, because these SOI effects are very small compared to the effects that we are  discussing below.

\section{Results and Discussion}
\subsection{defect structures}
\begin{figure}[tb]
\begin{center}
\includegraphics[width=15cm]{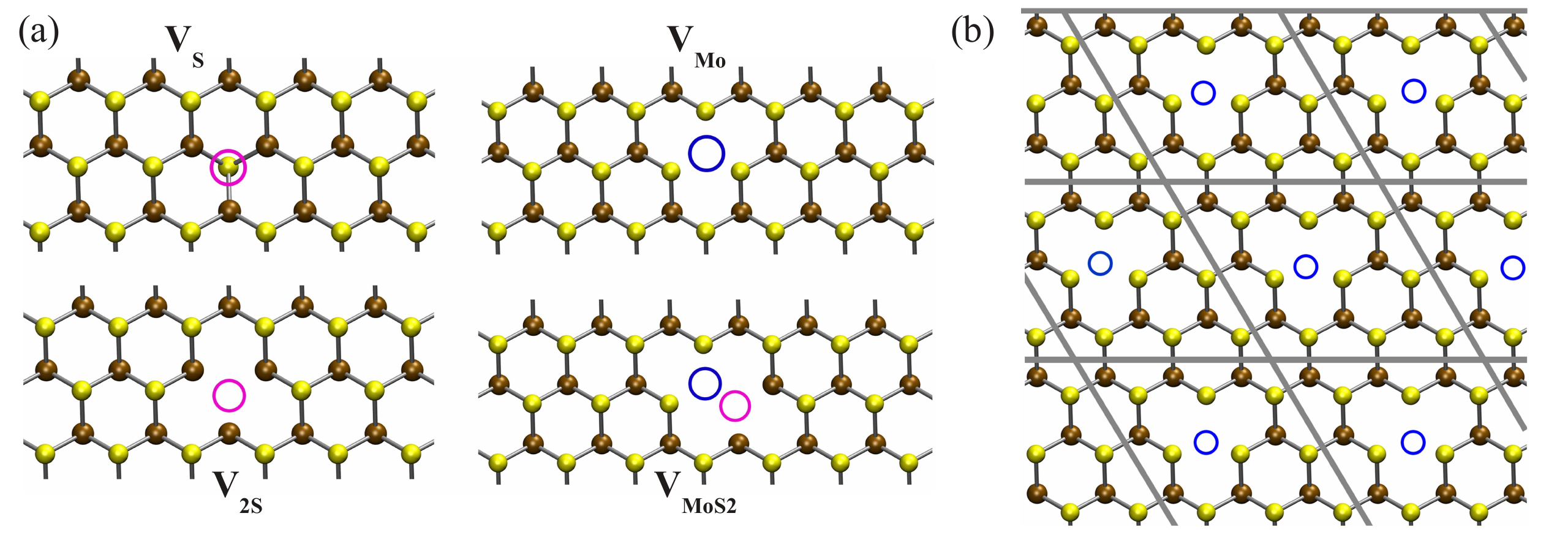}
\caption{
(a) The structures of the point defects models, considered in this study: single sulfur vacancy V$_S$, double sulfur vacancy V$_{2S}$, single molybdenum vacancy V$_{Mo}$, and MoS$_2$ vacancy V$_{MoS_2}$.
(b) shows the hexagonal supercells (gray lines) of monolayer MoS$_2$, used to simulate a defect concentration of 11.1\%.
In all panels brown and yellow balls represents molybdenum and sulfur atoms, respectively.
}
\label{fig:PD-struct}
\end{center}
\end{figure}

Figure \ref{fig:PD-struct}(a) shows the structure of the four vacancy point defects within ML-MoS$_2$ that are considered here.
The defect $V_S$ corresponds to a single S atom vacancy on one side of the ML, in $V_{2S}$ two S atoms are removed, one from the upper and one from the lower S layer, V$_{Mo}$ is a Mo atom vacancy and V$_{MoS_2}$ represents the absence of a MoS$_2$ unit.
Experimental \cite{Komsa2013b,Zhou2013,Hong2015} and theoretical studies \cite{Yuan2014a,Noh2014,Komsa2015,Haldar2015} of point defects highlight the importance of sulfur vacancy defects V$_S$ and V$_{2S}$, which are abundant because of their low defect formation energy. Therefore our work focuses on vacancy-type defects.
The point defects are placed in the center of hexagonal supercells of varying cell sizes, as shown in Fig.~\ref{fig:PD-struct}(b). So we study hexagonal arrangements of point defects of varying defect concentration. 
If the supercell contains one defect, then the defect concentration is $1/N_{cells}$, where $N_{cells}$ the number of primitive unit cells per supercell. Table \ref{tab:concentration} lists defect concentrations, defect separations and area densities of the considered systems. 

\begin{table}[tb]
\begin{tabular}{|l|r|r|r|r|}
\hline
{defect concentration (\%)} & 25.0 & 11.1 & 6.3 & 1.2 \\ \hline
{defect separation (\AA)} & 6.56 & 9.84 & 13.12 & 29.52 \\ \hline
{area density ($10^{13}$ defects/cm$^{-2}$)} & 26.83 & 11.93 & 6.708 & 1.325 \\ \hline
\end{tabular}
\caption{Parameters, characterizing MoS$_2$ monolayers with different concentrations of $V_S$, $V_{2S}$, V$_{Mo}$ and V$_{MoS_2}$ point defects.	
}
\label{tab:concentration}
\end{table}

\begin{figure}[tb]
\begin{center}
\leavevmode
\includegraphics[width=12cm]{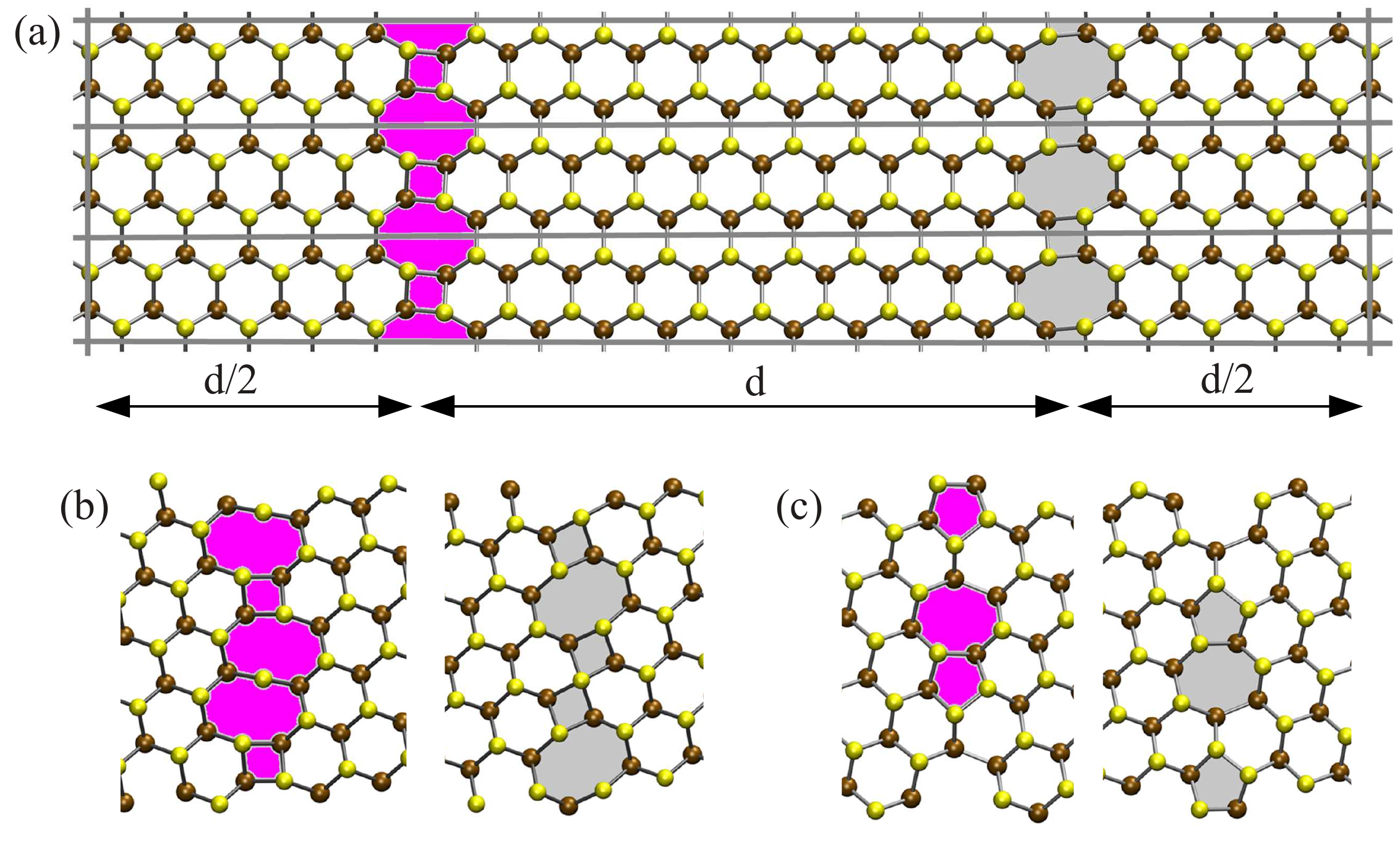}
\caption{
(a) The atomic structure and the unit cell (gray lines) of the 8-4 mirror grain boundary model in a MoS$_2$ monolayer. $d$ is the separation between the individual GBs.
Due to the periodic boundary conditions the simulated systems correspond to arrays of parallel, aligned  GBs. (b) 4-8-8 and 8-4-4 mirror GB structures and (c) 5-7 tilt GB structure. The left and right parts of the GBs are represented by magenta and gray color, respectively. Brown and yellow balls represents Mo and S-atoms, respectively.
}
\label{fig:LD-struc}
\end{center}
\end{figure}

The second type of systems, studied here, are periodic line defects or grain boundaries (GBs) as shown in Fig.~\ref{fig:LD-struc}.
We study three types of mirror grain boundaries, 8-4, 4-4-8 \cite{VanderZande2013} and 8-8-4 and one 5-7 tilt GB with a tilt angle of 38.2$^\circ$. 
The tilt angle is the misorientation angle of adjacent grains of ML-MoS$_2$ and the numbers (e.g. 8-4 or 5-7) correspond to the sequence of topological ``polygon'' defects that create the GB. Mind that the defects are only polygons in the planar projection and that their actual structure is three-dimensional.
The tilt angle of mirror GBs is 60$^\circ$ (or 180$^\circ$ or 300$^\circ$).
Furthermore there are different ways of constructing 5-7 line defects. Our 5-7 defects include Mo-Mo bonds in the 7-rings and S-S bonds in the 5-rings. They are structurally identical on the left and on the right side. Different types of 5-7 GB were previously studied. \cite{Zou2012}
Since we use periodic boundary conditions, the tilt, induced by one GB, has to be compensated by a second GB with an opposite tilt to create a periodic system. Therefore all our supercells contain two GBs, one on the left and one on the right. The supercells, containing  two GBs and the separation length $d$ between the GBs are illustrated in Fig.~\ref{fig:LD-struc}.

\subsection{\label{sec:localization} strongly localized defect states}

\begin{figure}[tb]
\begin{center}
\leavevmode
\includegraphics[width=15cm]{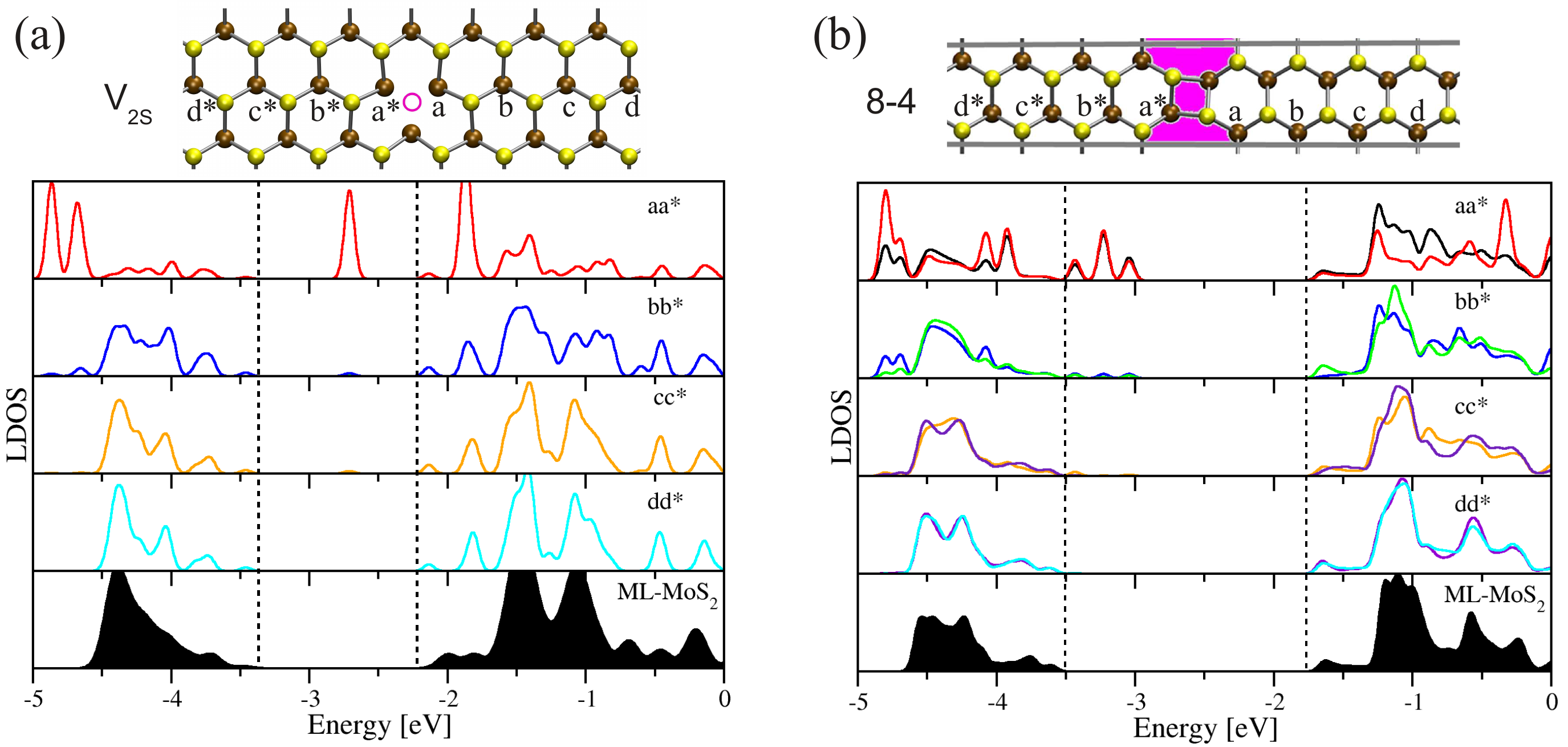}
\caption{
The strongly localized nature of (a) point and (b) line defect states. The defect state is localized on the shell of atoms directly surrounding the defect (a atoms), the second shell (b atoms) has very small contributions and atoms beyond that (c,d atoms) do not significantly contribute to the defect state.
(a) Panels, from top to bottom: the local density of states (LDOS) of the Mo atoms a,b,c,d near the V$_{2S}$ point defect (1.2\% defect concentration).  
(b) Similar results for the LDOS of the 8-4 grain boundary (GB separation is $d=67$ \AA). 
The last panel in (a) and (b) shows the total DOS of monolayer (ML) MoS$_2$.
} 
\label{fig:PD-localization}
\end{center}
\end{figure}

First, we discuss the observation that both point and line defect states in MoS$_2$ are strongly localized. This is demonstrated by the local density of states (LDOS) of Mo atoms along a line starting on the defect (a atoms) and moving away from it (atoms b,c,d) in Fig.\ref{fig:PD-localization}. The LDOS is the contribution of an individual atom to the total DOS. 
We analyzed the LDOS of all considered systems and the result was always as it is shown in Fig.\ref{fig:PD-localization} for the V$_{2S}$ point defect in (a) and the 8-4 GB in (b): the defect state is localized on the a atoms, i.e.~the shell of Mo atoms directly surrounding the point defect or, for GBs, the atoms involved in the ``polygon'' defect. The second shell of Mo atoms (b atoms) contributes very little to the defect state and atoms beyond that shell are basically bulk atoms with no significant contributions.

Another manifestation of the strong localization of the defect states is the relative insensitivity of the defect formation energy of point defects
on the defect concentration, even in the limit of high concentrations. 
We used DFTB to calculate the defect formation energy as described in the appendix and listed in table \ref{tab:Eform}. 
The relative increase of the defect formation energies of V$_S$ and V$_{2S}$ defects from concentrations of 1.2\% to 6.3\%,  11.1\%, and 25\% on the average is only 3\% and 7\%, 25\%, respectively. This result is particularly significant for systems with 25\% defects, where every fourth unit cell has a defect and still the defect formation energy for sulfur vacancies increases only by 25\% as compared to the dilute limit (1.2\%). 
The strongly localized nature of the defect states leads to very small interactions between individual defects, even if they are separated by only a few nanometers (also see the defect separations in Tab.~\ref{tab:concentration}).
This result explains why sulfur vacancy defects can be found in high concentrations. \cite{Qiu2013b} 
For simplicity, our defect structures are symmetrically arranged and homogeneously distributed. This is certainly an idealized situation. However, the strong localization of the defect states and the small electronic interactions between them, leads us to conjecture that the insights of this article will be largely independent of the specific arrangement of the point defects.

Similar calculations of the defect formation energy of the 8-4 GBs for different separations $d$ ranging between 13 and 67 {\AA} did not show any separation dependence. This shows the absence of long range stress fields and also agrees with the electronic structure analysis that shows that the defect states are strongly localized on a few shells of atoms near the actual defect and do not interact with other line defects that are nearby ($d > 13$ {\AA}).

\subsection{characterization of the in-gap states}

\begin{figure}[h]
\begin{center}
\leavevmode
\includegraphics[width=9cm]{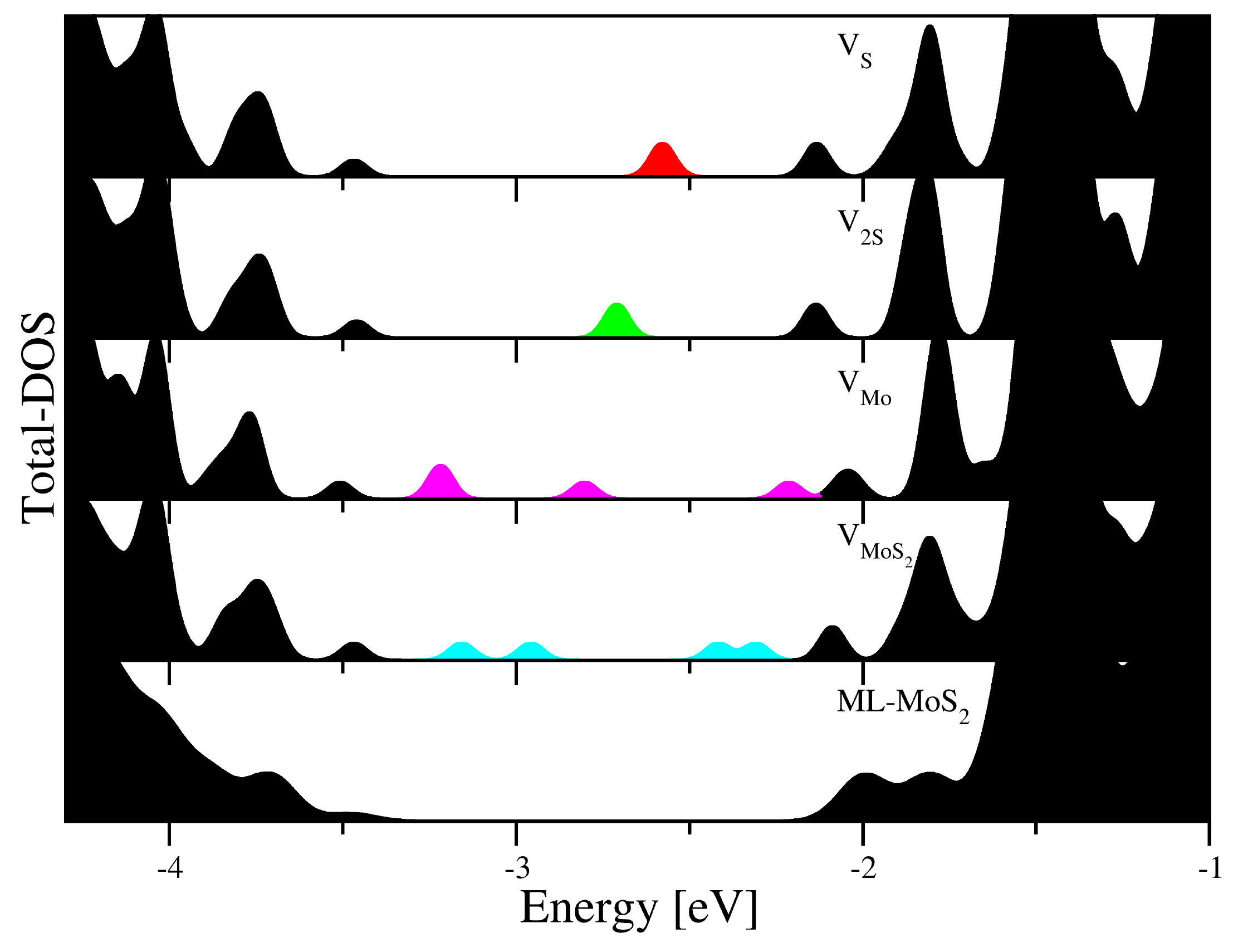}
\caption{
The 'fingerprints' of point defects: characteristic, defect induced in-gap states of the considered point defects shown by the total density of states (DOS). The in-gap states are highlighted in color and the bulk states (and ideal monolayer MoS$_2$) are shown in black.
}
\label{fig:in-gap-PD}
\end{center}
\end{figure}

Figure \ref{fig:in-gap-PD} indicates in color the in-gap states induced by different types of point defects. The highlighted states are 'deep levels'.
Additionally, the defects also induce 'shallow levels' that are close to the valence or condition band edges or levels that lie deep within these bands. In Fig.~\ref{fig:in-gap-PD} the latter states are not explicitly indicated. 

Single and double sulfur vacancies V$_{S}$, V$_{2S}$ are characterized by a single peak near the center of the band gap. The peak corresponds to two unoccupied states (this counting neglects the spin-degeneracy). \cite{Noh2014,Zhou2013,Pandey2016,Haldar2015} The difference between V$_{S}$ and V$_{2S}$ is that the level shifts to lower energies.
Mo point defects induce three deep levels. The two levels at lower energies are two-fold degenerate and the last one is a non-degenerate  state. \cite{Noh2014,Haldar2015}
Finally the V$_{MoS_2}$ defect has four levels in the gap. The second level (at about -3 eV) is two-fold degenerate the other three are single levels.
A comparision with full DFT calculations indicates that DFTB tends to place the in-gap states closer to the band edges, while full DFT places the levels more in the center of the gap. 
However, neither DFT nor DFTB are suitable to precisely determine the position of the in-gap states. This requires quasiparticle-based electronic structure methods that are computationally very demanding and go beyond the scope of the present work. 
%

The strong degree of localization combined with sharply peaked in-gap states could potentially explain the emission of single photons from point defects in ML WSe$_2$ (that is very similar to MoS$_2$). \cite{Srivastava2014a,He2015,Koperski2015,Chakraborty2015,Kumar2015}
Single photon emission is only possible if isolated quantum system with localized states are excited. The strong localization of the point defects creates  atom-like states with sharply peaked eigenstates. 
However electronic excitations between deep levels would result in photons with lower energies than the  absorption edge of the monolayer. But experimentally the frequency of the single-photons is observed at energies only slightly lower than this edge. Furthermore, electron-phonon coupling should broaden the localized in-gap states into vibronic bands as it is frequently observed in luminescent centers.\cite{Lounis2005a} The recent reports however found sharply peaked optical spectra. Further analysis about the nature of the single photon emission in TMDs is therefore necessary.


\begin{figure}[h]
\begin{center}
\leavevmode
\includegraphics[width=\columnwidth]{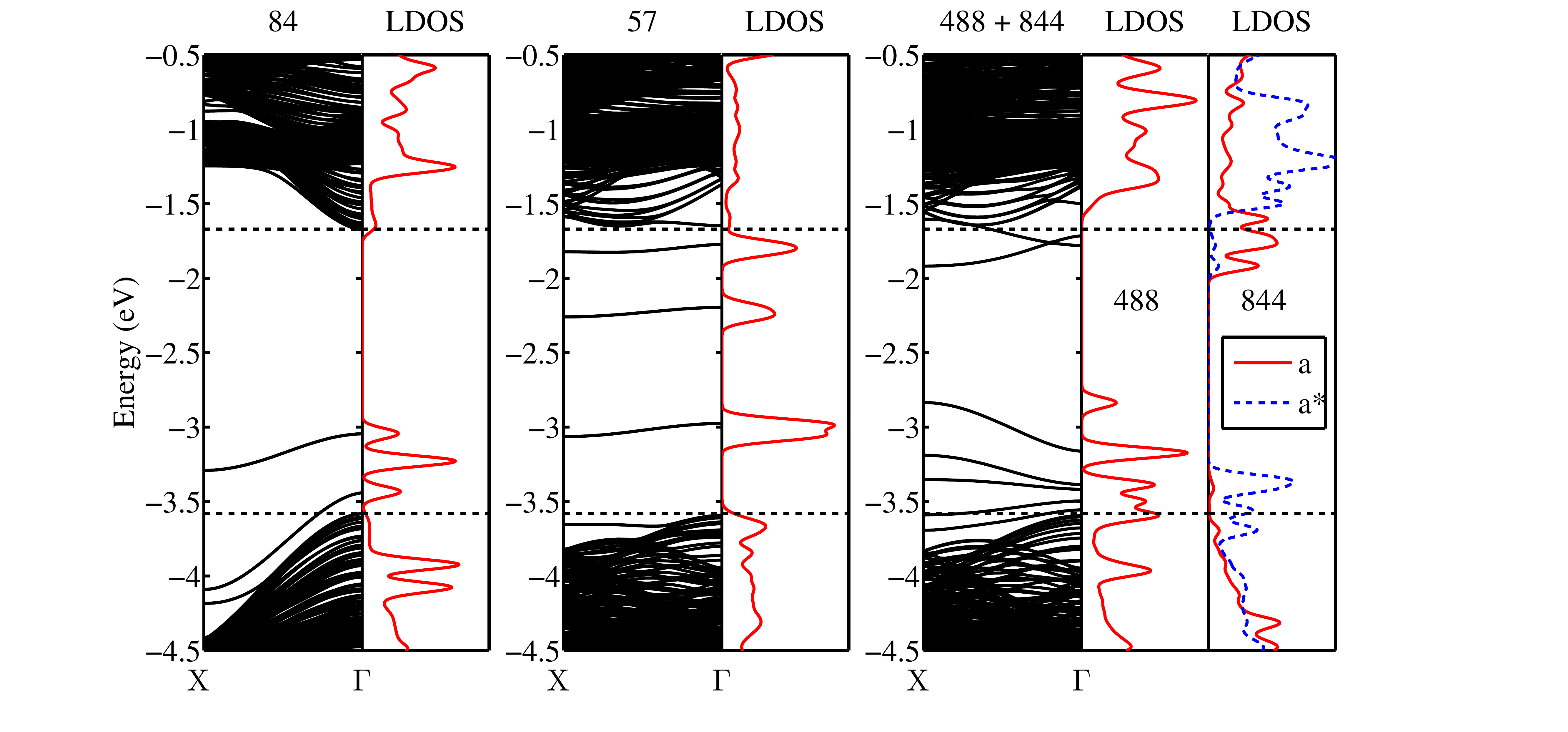}
\caption{
The 'fingerprints' of line defects: characteristic, defect induced in-gap states of the line defects shown by the band structure and the local density of states (LDOS) of a Mo atom at the defect. For the 488+844 system the LDOS is given for the individual line defects ($a$ is an atom at a 4-unit and a$^*$ is at a 8-unit of the 844 line defect). The monolayer band gap is indicated by dashed horizontal lines.
}
\label{fig:in-gap-LD}
\end{center}
\end{figure}

%

Also line defects induce in-gap states as shown for all considered GBs in Fig.~\ref{fig:in-gap-LD}. Here we consider periodic line defects as shown in Fig.~\ref{fig:LD-struc}. Such structures are actually an idealization. An analysis of real line defects in the transmission electron microscope reveal that real structures can rarely be considered as ideal and periodic.\cite{VanderZande2013} They are rather irregular with occasional periodic fragments. For the theoretical analysis such periodic fragments, however, offer a good model to analyze the basic change of the electronic structure due to GBs. 

The fact that our line defects are one-dimensional, periodic structures leads to the formation of a defect band structure that are related to states that are delocalized along the direction of the GB. However, as demonstrated in the previous section, the states do not spread out much into the direction perpendicular to the GB. The cosine-like dispersion of the defect bands leads to a broader energy signature in the density of states (Fig.~\ref{fig:in-gap-LD}) as compared to point defects. The most dispersive defect bands occur as typical double peaks in the DOS (e.g. 84-GB or 488-GB).

Similar to point defects also line defects create a characteristic 'fingerprint' of in-gap states.\cite{Zou2012,VanderZande2013,Zhou2013,Zou2015}
The 8-4 GB induces a single 'deep' defect band.\footnote{As our unit cells contain two GBs, the number of in-gap states is actually doubled.}
Its small dispersion creates two peaks in the DOS just below -3 eV in Fig.~\ref{fig:in-gap-LD}. The peak at about -3.5 eV corresponds to the onset of a shallow defect band. 
%
The  5-7 GB introduces 3 relatively dispersionless defect bands within the band gap. 
The band structure of the 488+844 system exhibits multiple defect bands where the distinction between shallow and deep levels is not always easy to make. 
The LDOS indicates which GB creates with states. The 488 GB generates the defect band around -3 eV (occuring as double-peak) and further levels above and below the valence band edge (VBE).
The states of the 844 GB can be distinguished between states related the structural 4-units ($a^*$ atoms) and the 8-units ($a$ atoms). The $a^*$ states are near the VBE and two bands near the conduction band edge (CBE) are related to $a$ atoms.
%
A comparison with DFT results shows again that DFTB places the in-gap states closer to the band edges.

\subsection{dependence of electronic structure on point defect concentration}
\begin{figure}[h]
\begin{center}
\leavevmode
\includegraphics[width=10cm]{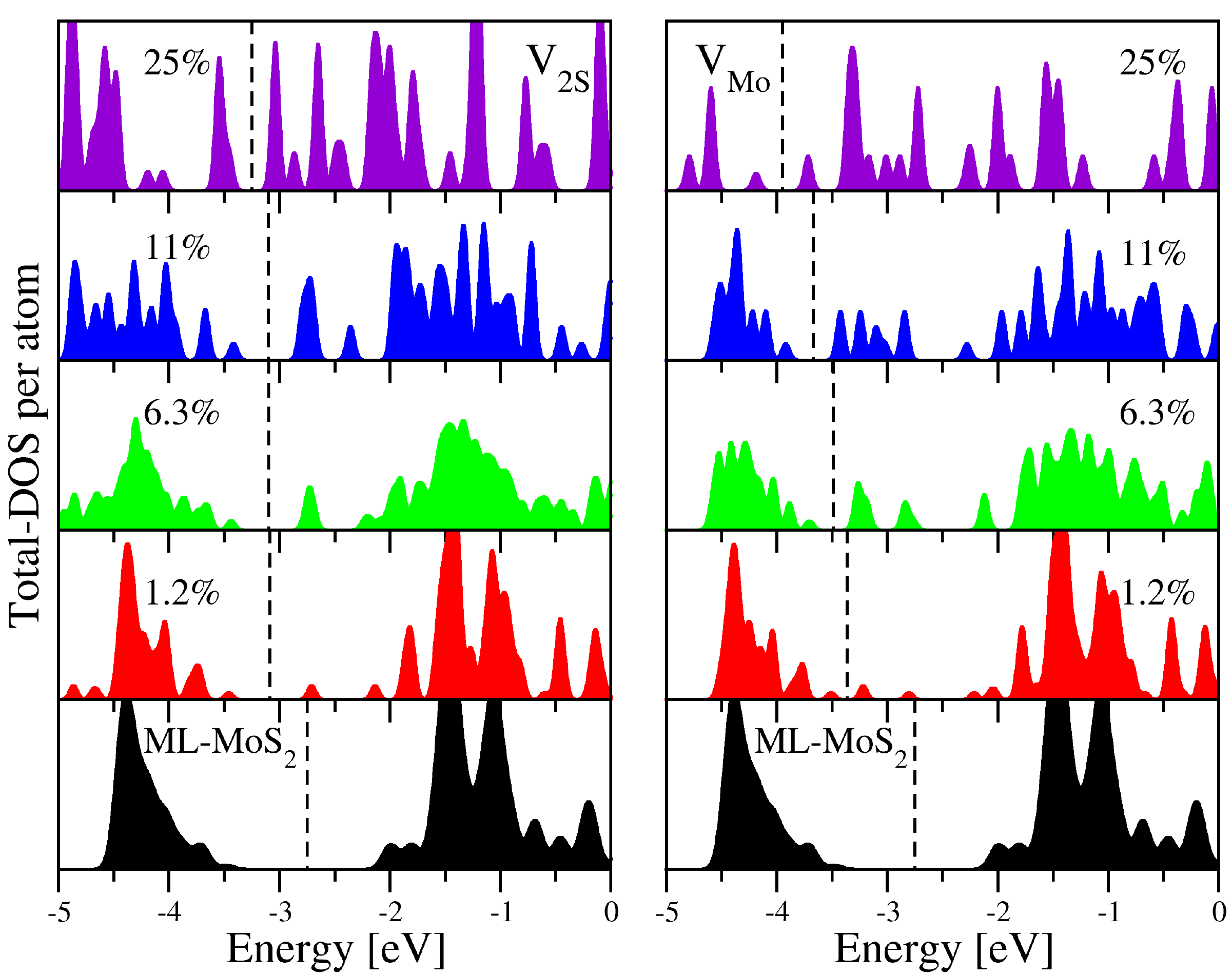}
\caption{
The electronic density of states (DOS) for different point defect models (V$_S$ and V$_{Mo}$) and point defect concentrations, varying between 1.2\% and 25.0\%, and the DOS of the pristine MoS$_2$ layer. The dashed lines represent the Fermi-energy. 
Point defects in concentrations of  up to 6\% (i) leave the bulk properties of MoS$_2$ nearly intact, (ii) create sharp in-gap states whose intensity merely increases with the concentration but their energy position is constant.
} 
\label{fig:PD-DOS}
\end{center}
\end{figure}

The influence of varying defect concentrations on the electronic structure is depicted in Fig.~\ref{fig:PD-DOS} for the examples of V$_{2S}$ and V$_{Mo}$. For defect concentration of up to 6\% the bulk electronic structure is left almost intact. With increasing concentration the in-gap states grow in intensity due to a stronger relative weight as compared to the ideal ML. Due to their strongly localized nature, defects do not interact much and therefore remain quasi-isolated objects. As a consequence the energy position of the in-gap states remains constant and is not changing as the concentration increases. Only for concentrations of 10\% and more, the modifications of the general electronic structure are significant and for 25\% even the bulk states are strongly influenced, as it to be expected for such high defect concentrations. The same trends were found for the other considered point defects (V$_S$ and V$_{MoS_2}$, not shown).

\subsection{impact on optical absorption}

\begin{figure}[h]
\begin{center}
\leavevmode
(a)\includegraphics[width=9cm]{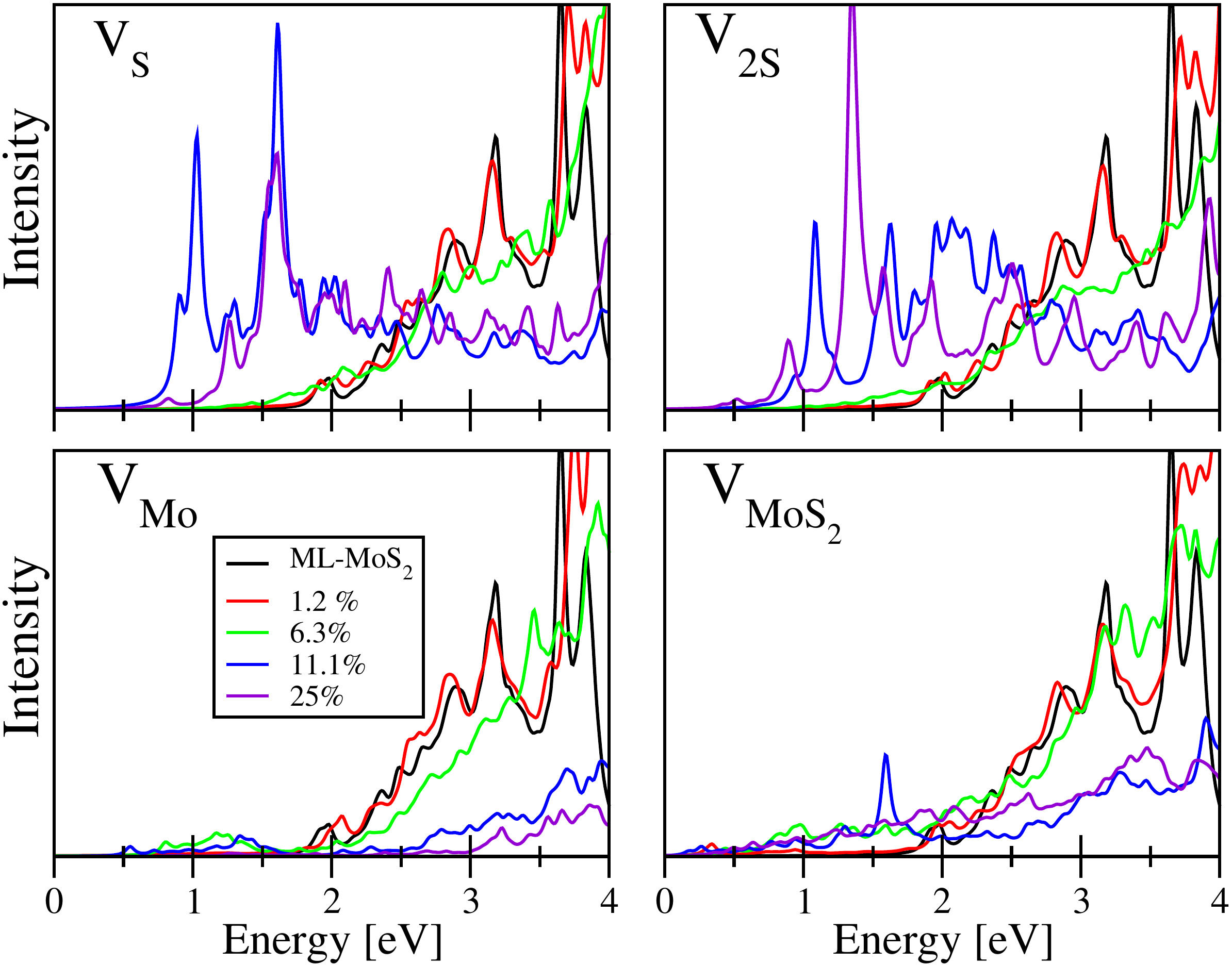}
(b) \includegraphics[width=4.5cm]{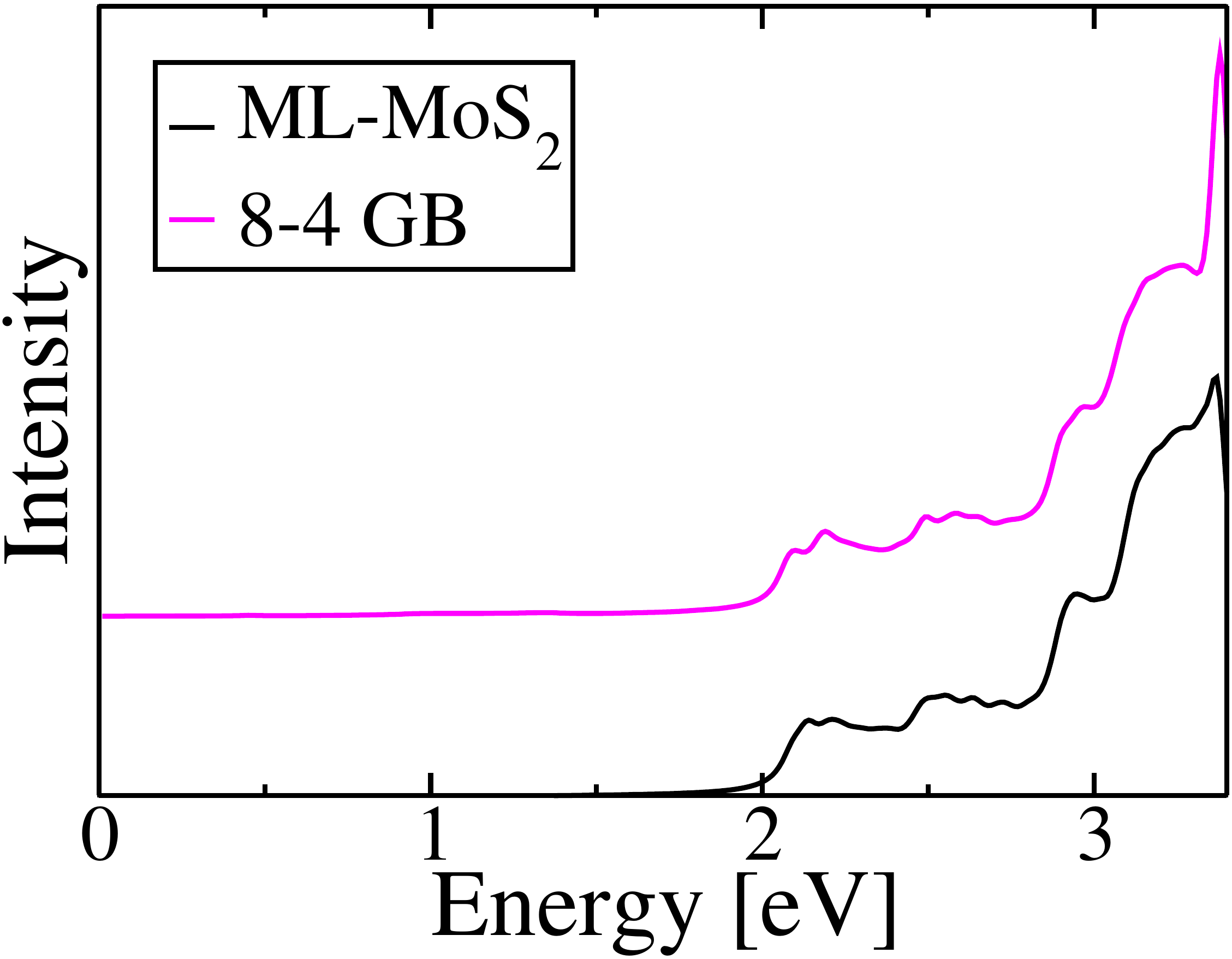}
\caption{
The impact point and line defects on the optical absorption spectrum of monolayer MoS$_2$. 
(a) The impact of the considered point defects for different concentrations. The defect-free system is shown as black line. For defect concentrations of 1.2\% (and below) the influence on the spectra is very small for all systems. 
(b) The impact of line defects on the absorption spectrum. As indicated by the comparison with the monolayer spectrum, the presence of the line defects does not have a significant influence on the optical absorption.
}
\label{fig:PD-absorpt}
\end{center}
\end{figure}

The concentration dependence of the optical and transport properties of random point defects in the limit of small concentration (1\% and smaller) was studied previously within an empirical tight-binding approach by Yuan et al.\cite{Yuan2014a} The general impact on the transport and optical properties at these concentrations was found to be rather small and our study confirms these results.
Here, we study the impact of point defects in the limit of higher defect concentrations (more than 1.2\%).  Figure \ref{fig:PD-absorpt}(a) shows our results which are based on the explicit calculations of the oscillator strength using TD-DFRT-TB which takes into account the symmetry of the orbitals, the dipole transition matrix elements as well-as many-body energy renormalizations of the transition energies. Such approach is necessary because the intensity of specific optical transitions cannot be inferred from pure electronic structure analysis as done in the previous section.
For the defect-free ML the ('bulk') absorption edge is found to be at 1.9 eV, which agrees well with experimental findings.\cite{Splendiani2010,Mak2010}
Increasing defect concentrations for all defects (i) lower the absorption at energies greater than 1.9 eV and (ii) introduce absorption peaks at energies smaller than 1.9 eV. 
Sulfur defects at high concentrations (above 10\%) introduce intense defect peaks, while for V$_{Mo}$ and V$_{MoS_2}$ the absorption is generally low. 
So the creation of V$_{Mo}$ defects in high concentrations would be a way to quench the the overall absorption in MoS$_2$, while samples with high concentrations of $V_{S}$ or $V_{2S}$ defects introduce intense spectral features below the bulk absorption edge.
However, at defect concentrations of 1.2\% and below the overall impact on the absorption spectrum for all type of defects is very small and the absorption spectrum nearly coincides with that of the defect-free ML.

Figure \ref{fig:PD-absorpt}(b) shows the absorption spectrum of the 8-4 GB. Although line defects introduce in-gap states (see Fig.~\ref{fig:in-gap-LD}) the related oscillator strength for the optical transitions is so small that it does not have a significant influence on the absorption and the spectrum is almost indistinguishable from the defect-free ML. For the other GB systems the result is the same (and therefore not explicitly shown).
We also checked the influence of the GB separation on the absorption but did not find any significant influence in the range ($d =32  \dots 67$ {\AA}). A comparison with table \ref{tab:concentration} shows that these are defects separations where in the case of point defects electronic interactions could be  neglected and our results for the GBs are consistent with this picture.

\section{Conclusions}
In view of recent experiments that explore physical phenomena of defective transition metal chalcogenides (doping, optical properties, single-photon emission) our study aimed at a deeper understanding of such systems. 
We employed electronic structure methods based on density functional theory to study the electronic and optical properties of single layer MoS$_2$ with different types of point and line defects.
For all types of defects we found that the electronic states that are induced by the defects are strongly localized on atoms that are forming the defect and on the shell of atoms surrounding the defect and do not significantly reach out much further in space. The strongly localized nature is further reflected in a weak dependence of the defect formation energy on the defect concentration or line defect separation. 
In the electronic energy spectrum the defect states occur as deep levels in the bulk band gap, as shallow levels very close to the band edges, as well as in-between the bulk states. Due to their strongly localized nature, all states of point defects are sharply peaked in energy. Periodic line defects form nearly dispersionless one-dimensional band structures and the related spectral features are also strongly  peaked.
Electronic structure analysis reveals that point defects in concentrations of up to 6\% leave the bulk properties of MoS$_2$ nearly intact and create sharp in-gap states whose intensity merely increases with the concentration but their energy position is constant.
The impact of point defects on the optical absorption for concentrations of 1\% and below is found to be very small. For higher defect concentrations molybdenum vacancies quench the overall absorption and sulfur defects lead to sharp absorption peaks below the absorption edge of the ideal ML. The considered line defects have such a low oscillator strength that there is practically no impact on the absorption spectrum.

\acknowledgments
This work was financially supported by International Max Planck Research School "Dynamical Processes in Atoms, Molecules and Solids" and the computational resources for this project were provided by 
ZIH Dresden. We thank Dr. Igor Baburin for fruitful discussions.

\appendix
\section{\label{sec:appendix}Defect Formation energy}
The defect formation energy as discussed in Sec.~\ref{sec:localization} is defined as
$$
E_\mathrm{form} = [E_\mathrm{system} - N_\mathrm{Mo} \mu_\mathrm{Mo} - N_\mathrm{S}  \mu_\mathrm{S}]/N_\mathrm{defect},
$$
where $E_\mathrm{system}$ is the total energy of a supercell with $N_\mathrm{defect}$ defects, $N_\mathrm{Mo,S}$ is the number of Mo or S atoms in the supercell, $\mu_\mathrm{Mo,S}$ is the chemical potential of Mo or S, respectively.
Using standard relations of thermodynamics this expression can be transformed to \cite{Groß2009}
$$
E_\mathrm{form} = [E_\mathrm{system} - (N_\mathrm{Mo} \ \mu_\mathrm{MoS_2}^\mathrm{bulk}) - 
(2 N_\mathrm{Mo} - N_\mathrm{S}) \mu_\mathrm{S}]/N_\mathrm{defect},
$$
where $\mu_\mathrm{MoS_2}^\mathrm{bulk}$ is the chemical potential of bulk MoS$_2$. The potential $\mu_\mathrm{S}$ varies between 
$$
\frac{1}{2} \Delta_R H + \mu_\mathrm{S}^\mathrm{bulk} \le \mu_\mathrm{S} \le \mu_\mathrm{S}^\mathrm{bulk},
$$
the Mo-rich limit (left-hand side) and the S-rich limit (right-hand side). $\Delta_R H$ is the heat of formation of MoS$_2$, and $\mu_\mathrm{S}^\mathrm{bulk}$ is the chemical potential of bulk sulfur (alpha-S).
To obtain the formation energies in table \ref{tab:Eform} the total energies $E_\mathrm{system}$ and $\mu_\mathrm{MoS_2}^\mathrm{bulk} = -176.191$ eV were calculated with DFTB, 
the parameters $\Delta_R H = -2.99$ eV, $\mu_\mathrm{S}^\mathrm{bulk} = -6.13$ eV, that merely define the range of the chemical potential $\mu_\mathrm{S}$, were adjusted to DFT values of $E_\mathrm{form}$ according to reference. \cite{Zhou2013}
The last line of table \ref{tab:Eform} provides a comparison of our DFTB formation energies with values obtained by DFT calculations. We obtain good agreement and the small deviation are within standard errors of the DFTB approximation.

\begin{table}[htbp]
\begin{tabular}{|r|r|r|r|r|}
\hline
 & \multicolumn{2}{c|}{V$_S$} & \multicolumn{2}{c|}{V$_{2S}$} \\ \hline
defect conc.(\%) & {Mo-rich} & {S-rich} & {Mo-rich} & {S-rich} \\ \hline
25.0 & 2.54 & 4.03 & 4.41 & 7.40 \\ \hline
11.1 & 2.11 & 3.61 & 3.61 & 6.61 \\ \hline
6.3 & 2.02 & 3.52 & 3.41 & 6.40 \\ \hline
1.2 & 1.97 & 3.46 & 3.26 & 6.26 \\ \hline
literature \cite{Zhou2013} (3 \%)& 1.5 & 3.0 & 3.0 & 6.0 \\ \hline
\end{tabular}
\caption{The concentration dependence of the formation energies of sulfur vacancy defects.}
\label{tab:Eform}
\end{table}

\bibliographystyle{nature} 
\bibliography{references}

\end{document}